\documentclass[12pt]{article}
\usepackage{epsfig}
\usepackage{amsfonts}
\begin {document}

\title {NEW REPRESENTATION OF SOME STATIC AND AXISYMMETRIC VACUUM SOLUTIONS}

\author{J.L. Hern\'andez-Pastora\thanks{E.T.S. Ingenier\'\i a
Industrial de B\'ejar. Phone: +34 923 408080 Ext
2263. Also at +34 923 294400 Ext 1527. e-mail address: jlhp@usal.es} $  ^{1,2}$ and J. Ospino$^1$ \\
\\
$^1$ Departamento de Matem\'atica Aplicada. \\ Universidad de Salamanca.  Salamanca,
Espa\~na.  \\
\\
$^2$ Instituto Universitario de F\'\i sica Fundamental y Matem\'aticas. \\ Universidad de Salamanca. Salamanca, Espa\~na}

\date{19 de Julio de 2010} \maketitle

\begin{abstract}
We solve the Einstein vacuum-equations for the case of static and axisymmetric solutions in a system of coordinates different from the Weyl standard one. We prove that there exists a  class of   solutions with the appropriate asymptotical behaviour which  can be written in a simple compact form, in terms of a function that must satisfies certain Cauchy-Newman problem. The relation between the choice of coordinates and the form of the metric functions that describe the solution is given by providing that analytic function which characterizes the metric as well as the gauge.
\end{abstract}


\newpage

\section{Introduction}

The general axially symmetric line element of an stationary space-time is commonly used to be written in isotropic coordinates, the so-called canonical Weyl coordinates \cite{kramer}. For the static case,  all the solutions with a good asymptotically behaviour are given, in this system of coordinates, by the Weyl family \cite{weyl}.

This choice of coordinates allow us to handle with only three metric functions to describe the kind of stationary solutions. The gauge imposes a determined value for the quantity $2\xi_{[\alpha}\eta_{\beta]}\xi^{\alpha}\eta^{\beta}$ involving the stationary ($\xi$) and axial ($\eta$) Killing vectors, which makes the length of both Killing vectors to depend on the same unique metric function. Nevertheless, there is no reason except for simplicity to perform the researching of stationary and axisymmetric solutions in that system of coordinates. In fact, there exist another representations of that kind of solutions, like the Erez-Rosen-Quevedo \cite{erq} and Gutsunayev-Manko \cite{gm} families, expressed in prolate spheroidal coordinates. In particular, the Weyl family of solutions is known to be specially ugly to describe the apparently more simple spherical solution. For this case, the Erez-Rosen-Quevedo representation arises as  one of the most suitable ways to write the Schwarzschild solution because one releases, in contrast with its Weyl form, the use of  a series to describe the corresponding metric functions. Indeed, the standard Schwarzschild coordinates are  a better adapted way to describe the spherical solution because, in addition to be a simpler expression of the metric function, the system of coordinates itself is related to the existence of a symmetry of the field equation which allow us to characterize the radial coordinate \cite{cqg2}.

There is a wide class of static and axisymmetric solutions with a very relevant physical characteristics for the exact description of the non-spherical space-time, the so-called {\it Pure Multipole Moments} \cite{mio}, which adopt a very cumbersome form in  standard Weyl coordinates. These are reasons for trying to look for static and axisymmetric solutions by solving the vacuum Einstein equations releasing the coordinate condition imposed by the use of Weyl canonical coordinates. As will be seen, the results obtained show the possibility of introducing from the beginning a new system of coordinates for the resolution of the static case. It is proved that the field equations can be solved in a non isotropic coordinate system, leading to solutions represented in terms of three metric functions (one more than the Weyl family). This solution adopt a very simple form than  its corresponding expression in the Weyl family, and the relation between them can be obtained by a well-defined change of coordinates. 

This work is organized as follows. In section 2 we make a brief review about the different descriptions of the general stationary and axisymmetric space-times. We want to put the attention on the fact that the Ernst equation for this kind of solutions still hold even for a different choice of coordinates since that equation represents an intrinsic description of the field equations.

In section 3 we solve the field equations for the static case providing  explicit expressions  for the two independent metric functions and  the other one derived by quadratures, and we discuss about the good asymptotical behaviour of the solution obtained.  In this section, a theorem is proved that allow to establish  a family of solutions depending on the election of determined analytic function that solves a Cauchy-Newman problem. In particular,  a series of  analytic functions fulfilling the conditions of the theorem is explicitly obtained, which leads to a concrete family of solutions of the field equations. Finally we show that the family of solutions obtained are of course included at the Weyl family but  they can be used to define  a different representation. The gauge of coordinates to change from one 
representation to another is explicitly shown in terms of the above-mentioned analytic function. 

A conclusion section contains some comments about the aims and results of the work as well as possible future extensions of it.

\section{The stationary and axisymmetric vacuum field equations}

Let $\nu_4$ be a Riemmanian manifold endowed with a stationary and axisymmetric metric $g_{\alpha \beta}$. Let $\xi$ and $\eta$ the corresponding Killing vectors associated to the time and axial isometries of this space-time  respectively.
As is known, there exist coordinates $\{x^{\beta}\}=\{t,\rho,z,\phi\}$ ($\beta=0..3$) adapted to both Killing vectors \cite{papa}, \cite{carter}, and
the more general stationary and axially symmetric line element can be written as follows:
\begin{equation}
ds^2=-e^{2\sigma}\left(dt^2-\omega d\phi\right)^2+e^{2\beta
-2\sigma}\left(dz^2+d \rho^2\right
)+J^2 e^{-2\sigma}d\phi^2 \ ,\label{metric}
\end{equation}
where $\sigma$, $\beta$, $\omega$ and $J$ are metric functions depending on $x^1\equiv \rho$ and $x^2\equiv z$.
The Einstein's vacuum field equations for this metric
(\ref{metric}) are the following:

\begin{equation}
\sigma ^{\prime \prime}+\ddot \sigma+\frac{\dot J}{J}\dot
\sigma+\frac{J^\prime}{J}\sigma^\prime+\frac{1}{2}\frac{e^{4\sigma}}{J^2}\left ((\omega^\prime)^2+\dot \omega ^2\right )=0\label{1}
\end{equation}

\begin{equation}
\omega ^{\prime \prime}+\ddot \omega+4\omega^\prime\sigma^\prime+4\dot \omega \dot
\sigma-\frac{J^\prime}{J}\omega ^\prime-\frac{\dot J}{J}\dot\omega=0\label{2}
\end{equation}

\begin{equation}
(\sigma ^\prime)^2+\dot \sigma^2+\frac{1}{4}\frac{e^{4\sigma}}{J^2}\left (\dot \omega^2+(\omega^\prime)^2\right )+\beta ^{\prime \prime}+\ddot \beta=0
\label{3}
\end{equation}

\begin{equation}
2\dot \sigma \sigma ^\prime+\frac{\dot
J^\prime}{J}-\frac{1}{2}\frac{e^{4\sigma}}{J^2}\dot \omega \omega^\prime=\frac{J^\prime}{J}\dot \beta+\frac{\dot J}{J}\beta^\prime\label{4}
\end{equation}

\begin{equation}
(\sigma ^\prime)^2-\dot \sigma ^2+\frac{J^{\prime\prime}}{J}+\frac{1}{4}\frac{e^{4\sigma}}{J^2}\left(\dot
\omega^2-(\omega ^\prime)^2\right)=\frac{J^\prime}{J}\beta^\prime-\frac{\dot J}{J}\dot \beta \label{5}
\end{equation}

\begin{equation}
\ddot J+J^{\prime\prime}=0 \ ,
\label{6}
\end{equation}
where  $\prime$ and  $\dot{}$ denote  derivatives with respect to the coordinate $\rho$ and $z$ respectively. The metric functions $\sigma$, $\omega$ and $J$ can be characterized intrinsically from the Killing vectors by the following scalars:
\begin{eqnarray}
\xi^{\alpha}\xi_{\alpha}&=&e^{-2\sigma} \quad , \quad \xi^{\alpha}\eta_{\alpha}=e^{2\sigma}\omega \nonumber\\
\eta^{\alpha}\eta_{\alpha}&=&e^{-2\sigma}J^2-e^{2\sigma}\omega^2 \quad , \quad 2\xi_{[\alpha}\eta_{\beta]}\xi^{\alpha}\eta^{\beta}=J^2 \ .
\label{scalars}
\end{eqnarray}

An alternative intrinsic and compact form of writing the line element (\ref{metric}) is the following
\begin{equation}
 ds^2=-f(dx^0-\varphi_i dx^i)^2+\hat g_{ij} dx^i dx^j ,
\end{equation}
$\hat g_{ij}$ being the quotient metric on the manifold $\nu_3$, and $f\equiv e^{2\sigma}$, $g_{0i}=f\varphi_i$, $g_{0i}=0, i\neq3$, $\varphi_3 \equiv \omega$. Bel \cite{bel} and Geroch \cite{geroch} proposed independently to write the Einstein equations by using the conformal metric $\bar g_{ij}=f\hat g_{ij}$, as follows
\begin{eqnarray}
& f \bar{\bigtriangleup}_2f-\bar{\bigtriangleup}_1f+\bar{\bigtriangleup}_1W=0\nonumber\\
& f \bar{\bigtriangleup}_2W-2\bar{\bigtriangleup}_1(f,W)=0 \nonumber\\
&\bar R_{ij}=\frac 12 f^{-2}\left(\bar{\nabla}_iW\bar{\nabla}_jW+\bar{\nabla}_if\bar{\nabla}_jf\right)\ ,
\end{eqnarray}
where $\bar R_{ij}$ denotes the Ricci tensor associated to the conformal metric $\bar g_{ij}$, $\bar{\bigtriangleup}_1(A,B)\equiv \bar g^{ij} \partial_i A \partial_j B$, $\bar{\bigtriangleup}_2(A)\equiv \bar g^{ij}\bar{\nabla}_i\bar{\nabla}_j A$ represent the Beltrami operators, and the scalar $W$ is defined from the vorticity $\Omega_{ij}=f^{1/2}\left(\partial_i\varphi_j-\partial_j\varphi_i\right)$ as follows\footnote{Let us remind that the existence of the scalar $W$ is derived from  one of the Einstein vacuum equation (\ref{2}), $\hat\nabla_i\hat\Omega^{ij}-2\Lambda_i\hat\Omega^{ij}=0$, in terms of the so-called {\it gravitational fields} $\Lambda_i$, $\Omega_{ij}$ (aceleration and vorticity)}
\begin{equation}
 \hat\Omega_j=f^{-1}\partial_jW ,
\end{equation}
where $\hat \Omega_j=\hat g_{jk}\frac 12 \hat \epsilon^{kil}\Omega_{il}$, and the derivatives of $W$ for the case of our metric (\ref{metric}) are $\dot W=f^2J^{-1}\omega^{\prime}$, $W^{\prime}=-f^2J^{-1}\dot \omega$.

Finally, Geroch \cite{geroch} introduced the complex function $E=f+iW$, the {\it Ernst potential} (\ref{ernst}), in terms of which the above shown Einstein equations (\ref{1})-(\ref{6}) can be written as follows
\begin{eqnarray}
& & \left(E+E^{\star}\right)\bar{\bigtriangleup}_2E-2\bar{\bigtriangleup}_1E=0 \nonumber\\
& & \bar R_{ij}=\left(E+E^{\star}\right)^{-2}\left(\bar{\nabla}_iE\bar\nabla_jE^{\star}+\bar{\nabla}_jE\bar\nabla_iE^{\star}\right) \ .
\label{ernst}
\end{eqnarray}

\section{The static case}

\noindent If we suppose that $\omega=0$,   the
system of equations (\ref{1})-(\ref{6}) is  reduced to:
\begin{equation}
\sigma ^{\prime \prime}+\ddot \sigma+\frac{\dot J}{J}\dot
\sigma+\frac{J^\prime}{J}\sigma ^\prime=0\label{W1}
\end{equation}
\begin{equation}
(\sigma ^\prime)^2+\dot \sigma ^2+\beta ^{\prime \prime}+\ddot
\beta=0 \label{W2}
\end{equation}
\begin{equation}
2\dot \sigma \sigma ^\prime+\frac{\dot
J^\prime}{J}=\frac{J^\prime}{J}\dot \beta+\frac{\dot J}{J}\beta
^\prime\label{W3}
\end{equation}
\begin{equation}
(\sigma ^\prime)^2-\dot \sigma ^2+\frac{J^{\prime
\prime}}{J}=\frac{J^\prime}{J}\beta ^\prime-\frac{\dot J}{J}\dot
\beta \label{W4}
\end{equation}
\begin{equation}
\ddot J+ J^{\prime \prime}=0\label{W5} \ .
\end{equation}

\subsection{A solution of the equations}

We proceed now to rewrite the above equations by using the following complex variables:
\begin{equation}
u=\rho+iz \quad , \quad \bar u=\rho-iz \quad ,
\end{equation}
and the equations (\ref{W1}) and (\ref{W5}) turn out to
be the following:
\begin{equation}
2J\sigma _{u\bar u}+J_{u}\sigma _{\bar u}+J_{\bar u}\sigma _{u}=0
\label{C2}
\end{equation}
\begin{equation}
J_{\bar u u}=0\label{C3} \ .
\end{equation}

\noindent A general solution of the equation (\ref{C3}) is given
by
\begin{equation}
J(u, \bar u)=J_1(u)+J_2(\bar u) \ ,
\label{jota}
\end{equation}
and therefore, the equations for the metric functions $\sigma$ and
$\beta$ can be solved, in separated variables, as follows:
\begin{equation}
\displaystyle{\sigma(u,\bar u)=\frac{c}{\sqrt{(aJ_1(u)+b)(aJ_2(\bar u)-b)}}}
\label{sigma}
\end{equation}

\begin{equation}
\displaystyle{\beta(u,\bar u)=-\frac 18 a^2J^2\sigma^4+\frac 12 \ln\left(J_u J_{\bar u}\right)+\alpha} \ ,
\label{beta}
\end{equation}
where $\alpha,a,b,c$ are  arbitrary constants and the subindices denote derivation with respect to the corresponding variable.

\subsection{The asymptotical conditions}

Let us now consider the appropriated behaviour of these solutions; since we want to describe the gravitational field of isolated compact bodies, the metric is required to be  asymptotically flat, i.e.,  in a neighborhood of infinity the line element must resembles the Minkowski metric
\begin{equation}
ds^2=-dt^2+d\rho^2+dz^2+\rho^2 d \varphi^2 \quad ,
\end{equation}
and therefore, the metric functions must fulfill the following asymptotical conditions:
\begin{equation}
\lim_{\rho\rightarrow \infty}  J= \rho \label{condijota}
\end{equation}
\begin{equation}
\lim_{\rho\rightarrow \infty} \sigma = 0  \label{condisigma}
\end{equation}
\begin{equation}
\lim_{\rho\rightarrow \infty} \beta = 0 \label{condibeta} \ .
\end{equation}

Furthermore, a {\it regularity condition} is required on the $\varphi$ coordinate to have the standard periodicity $2 \pi$ \cite{kramer}:
\begin{equation}
\lim_{\rho\rightarrow 0}\left[\frac{l_{\mu}l^{\mu}}{4 l} \right]=1 \ ,
\label{rc}
\end{equation}
where $l\equiv g_{33}=e^{-2\sigma}J^2$ denotes the length of the spacelike Killing vector $\partial_{\varphi}$ that represents the axial symmetry. This regularity condition for the solution (\ref{jota}) leads to the following limit:
\begin{equation}
\lim_{\rho \rightarrow 0}\left(J_{\rho}^2+J_z^2\right)e^{-2 \beta} =1
\label{rc2}
\end{equation}
if the following additional condition holds:
\begin{equation}
\lim_{\rho \rightarrow 0}J=\rho=0 \label{otramas} \ .
\end{equation}

Let us choose the functions $J_1(u)$ and $J_2(\bar u)$ in such a way that the function $J(u,\bar u)$,
\begin{equation}
J(u,\bar u)=\frac 12 u+f_1(u)+\frac 12 \bar u+f_2(\bar u)\label{efes}
\end{equation}
fulfills the asymptotical condition (\ref{condijota}) iff the arbitrary functions $f_1(u)$ and $f_2(\bar u)$ goes to zero a infinity. Since we look for a real function $J$, the imaginary terms  of the complex functions $f_1(u)$ and $f_2(\bar u)$ must be equal except for their signs, and hence we shall consider these functions to be complex conjugated as follows:
\begin{equation}
f_1(u)=v(\rho,z)+i w(\rho,z) \quad , \quad f_2(\bar u)=v(\rho,z)-i w(\rho,z) \label{efes2} \ ,
\end{equation}
where $v(\rho,z)$ and $w(\rho,z)$ are harmonic real functions that verify the Cauchy-Riemman conditions,i.e,
\begin{equation}
v_{\rho \rho}+v_{zz}=0 \quad, \quad v_{\rho}=w_{z} \quad , \quad v_z=-w_{\rho}
\label{cr}
\end{equation}
since $J(u,\bar u)$ (as well as $f_1(u)$) is a holomorphic function, solution of the equation (\ref{C3}).

Hence, the asymptotical conditions (\ref{condijota}),(\ref{condisigma}) are fulfilled whenever
\begin{equation}
\displaystyle{\lim_{\rho\rightarrow\infty}v=0} ,\label{uvecero}
\end{equation}
and the condition (\ref{condibeta}) for the metric function $\beta$ requires that
\begin{equation}
\lim_{\rho\rightarrow \infty} \left(J_{\rho}^2+J_z^2\right)=4e^{-2\alpha} \label{infini} \ .
\end{equation}
Finally, with respect to the regularity condition, since we impose to the function $v(\rho, z)$ to be zero at the symmetry axis (\ref{otramas}), then the equation (\ref{rc}) leads to the following limit:
\begin{equation}
\lim_{\rho\rightarrow 0} 4 e^{-2 \alpha} =1
\label{zero}
\end{equation}
since $\displaystyle{J_uJ_{\bar u}=\frac 14 (J_{\rho}^2+J_z^2)}$. Therefore, this condition (\ref{zero}) implies that $\alpha=\ln 2$ and the equation (\ref{infini}) turns out to be
\begin{equation}
\lim_{\rho \rightarrow\infty}\left(J_{\rho}^2+J_z^2\right)=1 \label{infini2}.
\end{equation}

Hence, we can hold the following theorem:

\vspace{5mm}

\noindent {\bf Theorem 1}

 The following metric functions:
 \begin{eqnarray}
 &J(\rho,z)=\rho+2v(\rho,z) \nonumber\\
 &\displaystyle{\sigma(\rho,z)=\frac{c}{\sqrt{(aJ_1+b)(aJ_2-b)}}}\nonumber\\
 &\displaystyle{\beta(\rho,z)=-\frac 18 a^2J^2\sigma^4+\frac 12 \ln\left(J_{\rho}^2+J_z^2\right)} \ ,
\label{teorem}
\end{eqnarray}
where $J_1(\rho,z)=\displaystyle{\frac12(\rho+iz)+v+iw}$, $J_2(\rho,z)=\displaystyle{\frac12(\rho-iz)+v-iw}$ and $a$, $b$, $c$ are arbitrary constants, determine a static and axisymmetric  vacuum solution of the Einstein equations, with the appropriate asymptotical and boundary behaviour, if the function $v(\rho,z)$ is a solution of the following Cauchy-Newman problem:
\begin{eqnarray}
v_{\rho\rho}+v_{zz}=0 \quad , \quad &v(\rho,z)\big|_{\rho\rightarrow 0,\infty}=0 \nonumber\\
&v_{\rho}(\rho,z)\big|_{\rho\rightarrow \infty}=0 \nonumber\\
&v_{z}(\rho,z)\big|_{\rho\rightarrow \infty}=0 \ ,
\label{cauchynewman}
\end{eqnarray}
with $w(\rho,z)$ being a solution of the Cauchy-Riemman conditions: $w_z=v_{\rho}$, $w_{\rho}=-v_z$.

\vspace{2mm}

\noindent {\bf Proof}: The asymptotical conditions (\ref{condijota}-\ref{condibeta}) and the regularity condition (\ref{rc}) are fulfilled by the metric functions (\ref{jota}-\ref{beta}) if we use the equations (\ref{efes}-\ref{cr}) and we choose a function $v(\rho,z)$ that satisfies the condition (\ref{uvecero}) and goes to zero at the symmetry axis. Since $J_{\rho}^2+J_z^2=1+4(v_{\rho}+v_{\rho}^2+v_z^2)$, the conditions (\ref{cauchynewman}) lead to verify the equation (\ref{infini2}) and it allows us to conclude the proof.

\subsection{A family of solutions}

A solution of the Cauchy-Newman problem (\ref{cauchynewman}) is given by the following series:
\begin{equation}
\displaystyle{ v(\rho,z)=\sum_{n=0}^{\infty}\frac{a_{2n+1}}{r^{2n+1}}\cos\left[(2n+1)\theta_c\right]+
\frac{b_{2n}}{r^{2n}}\sin\left[(2n)\theta_c\right]} \label{familia} \ ,
\end{equation}
where $r\equiv\sqrt{\rho^2+z^2}$ and $\theta_c$ being the complementary angle $\displaystyle{\theta_c\equiv\frac{\pi}{2}-\theta}$ of the standard polar angle $\theta$, and  defined by\footnote{Let us note that the angle $\delta$ is exactly equal to $\theta_c$ since $\cot\theta=\tan\theta_c$} $\displaystyle{\theta_c=\delta\equiv\arctan \frac{z}{\rho}}$.

This family of functions is obtained from the following considerations: the successive negative powers of the complex variable $u$ provide a series of holomorphic functions $F_k(u)=u^{-k}$ solutions of the harmonic equation (\ref{C3}), and therefore, their real and imaginary parts are real functions fulfilling the  equation (\ref{W5}) in the variables $(\rho,z)$. And furthermore, if we take alternatively the real part of the functions $F_k(u)$ for even $k$ and the imaginary part of those functions with odd order $k$, we  get a set of real functions which verify the Cauchy-Newman conditions (\ref{cauchynewman}).

Therefore, according to (\ref{teorem}) and (\ref{familia}), the metric function $J(\rho,z)$ can be written as follows:
\begin{equation}
\displaystyle{ J(\rho,z)=\rho\left[1+2\sum_{n=0}^{\infty}\frac{a_{2n+1}}{r^{2n+2}}(-1)^n\frac{\sin\left((2n+1)\theta\right)}{\sin\theta}+
\frac{b_{2n}}{r^{2n+1}}(-1)^{n+1}\frac{\sin\left((2n)\theta\right)}{\sin\theta}\right]} \label{jota1} \ .
\end{equation}

And, by developing the trigonometric relations we can write the following expression:
\begin{equation}
\displaystyle{ J(\rho,z)=\rho\left[1+2\sum_{n=0}^{\infty}\frac{a_{2n+1}}{r^{2n+2}}C^{(1)}_{2n}(\Omega)+
\frac{b_{2n}}{r^{2n+1}}C^{(1)}_{2n-1}(\Omega)\right]} \label{jota2} \ ,
\end{equation}
with $\Omega\equiv\cos\theta$, and $C^{(1)}_n(\Omega)$ are the Gegenbauer polynomials of degree $n$, since we know that
\begin{eqnarray}
&\cos\left[(2n+1)\theta_c\right]=(-1)^n\sin\left[(2n+1)\theta\right]= C^{(1)}_{2n}(\Omega) \sin\theta\nonumber\\
&\sin\left[(2n)\theta_c\right]=(-1)^{n+1}\sin\left[(2n)\theta\right]= C^{(1)}_{2n-1}(\Omega) \sin\theta\label{gegen} \ .
\end{eqnarray}

Therefore, accordingly  to equations (\ref{gegen}) we can write the function $v(\rho,z)$ as follows:
\begin{equation}
v(\rho,z)=\sin\theta \sum_{n=0}^{\infty}\frac{h_n}{r^{n+1}}C^{(1)}_n(\Omega) \label{lauve} \ ,
\end{equation}
for any set of arbitrary parameters $h_n$, and the metric function $J$ as follows\footnote{In fact, it is easy to verify that this expression is the general solution, in separated variables, of the equation (\ref{W5}).}:
\begin{equation}
J(\rho,z)=\rho \left[1+2\sum_{n=0}^{\infty}\frac{h_n}{r^{n+2}}C^{(1)}_n(\Omega)\right] \label{lajota} \ .
\end{equation}

From the equation (\ref{teorem}) we have that  the  metric function $\sigma(\rho,z)$ is given by the following expression:
\begin{equation}
\sigma(\rho,z)=c \left[a^2(\frac14 r^2+v^2+w^2+\rho v+z w-\frac{b^2}{a^2})-iab(z+2w)\right]^{-1/2} \ , \label{sigma1}
\end{equation}
which  can be simplified as follows:
\begin{equation}
 \sigma(\rho,z)=\frac{2c}{a}\left[J^2+(z+2w)^2-\left(\frac {2b}{a}\right)^2-i\frac{4b}{a}(z+2w)\right]^{-1/2}
 \label{sigma2} .
\end{equation}
The function $w(\rho,z)$ is obtained by integration of a quadrature from the equation (\ref{familia}), and this is the result:
\begin{equation}
w(\rho,z)=2\sum_{n=0}^{\infty}\frac{h_n}{r^{n+1}}\left[C^{(1)}_{n-1}(\Omega)-\Omega C^{(1)}_n(\Omega)\right].
\end{equation}
Finally, the metric function $\beta(\rho,z)$ is given by the following expression:
\begin{equation}
\displaystyle{\beta(\rho,z)=-2\frac{c^4}{a^2}\frac{J^2}{\left[J^2+(z+2w)^2\right]^2}+\frac 12\ln\left[1+4(v_{\rho}+v_{\rho}^2+v_z^2)\right]} \label{beta1} \ ,
\end{equation}
where the derivatives of the function $v(\rho,z)$ with respect to $\rho$ and $z$ are given by the following expressions:
\begin{eqnarray}
 &\displaystyle{v_{\rho}=\sum_{n=0}^{\infty}\frac{h_n
 (n+1)}{r^{n+2}}\left[(2\Omega^2-1)C^{(1)}_n(\Omega)-\Omega
 C^{(1)}_{n-1}(\Omega)\right]} \nonumber\\
&\displaystyle{v_z=\sin\theta \sum_{n=0}^{\infty}\frac{h_n(n+1)}{r^{n+2}}\left[-2\Omega C^{(1)}_n(\Omega)+C^{(1)}_{n-1}(\Omega)\right]} \label{derilauve} \ .
\end{eqnarray}

\subsection{The Weyl limit}

The Weyl family of solutions  \cite{weyl} are written in a special system of isotropic coordinates $\{\hat \rho, \hat z\}$ namely the canonical Weyl coordinates. This choice of coordinates makes able to fix the metric function $J$ of the general line element (\ref{metric}) in such a way that it equals  the Weyl coordinate $\hat \rho$, Hence, the problem of searching for solutions is reduced to solve only one differential equation for the metric function $\sigma$ (\ref{W1}).

If we impose this condition on the metric function $J$, i.e., we take  $J(\rho,z)=\rho$, we are equivalently considering  a special choice of coordinates and $\{\rho,z\}$ become the canonical Weyl coordinates. Therefore, by taking $v(\rho,z)=0$, and consequently
\begin{equation}
J_1=\frac12 u \quad , \quad J_2=\frac 12 \bar u ,
\end{equation}
we have that
\begin{equation}
\sigma(u,\bar u)=\frac{2c}{\sqrt{(au+2b)(a\bar u-2b)}}=\frac{2c}{a r\sqrt{1-2 \Omega \lambda +\lambda^2}}\label{unasigma} \ ,
\end{equation}
where $\displaystyle{\lambda\equiv \frac{2bi}{a r}}$. By taking into account the generator function of the Legendre polynomials $P_n(\Omega)$ we can conclude that
\begin{equation}
\sigma(r,\Omega)=\frac {2c}{a} \sum_{n=0}^{\infty}\frac{(i 2b/a)^n}{r^{n+1}}P_n(\Omega) \label{weyl} \ ,
\end{equation}
and the metric function $\beta(\rho,z)$ is given by the following expression:
\begin{equation}
\beta(\rho,z)=-\frac 18 a^2 \rho^2 \sigma^4 \ .
\end{equation}

In conclusion, these metric functions represent a family of static  and axisymmetric vacuum solutions that verify the following asymptotical conditions:
\begin{equation}
\lim_{\rho\rightarrow \infty}  \beta= 0 \quad , \quad \lim_{\rho\rightarrow \infty} \sigma = 0 , \label{weylsigmaybeta}
\end{equation}
and the regularity condition for this case, i.e.,
\begin{equation}
\lim_{\rho\rightarrow 0} \beta = 0 . \label{weylbeta}
\end{equation}

The  coefficients $a^W_n$ of this solution (\ref{weyl}) in the Weyl representation\footnote{All the solutions are described by the following metric function: $\displaystyle{\sigma(\hat \rho,\hat z)=\sum_{n=0}^{\infty}\frac{a_n^W}{\hat r^{n+1}}P_n(\hat z / \hat r)}$, with $\hat r \equiv \left(\hat \rho^2+\hat z^2\right)^{1/2}$}  are given by the expression $\displaystyle{a^W_n\equiv\frac{2c}{a}\left(\frac{i2b}{a}\right)^n}$, and these particular coefficients are not arbitrary because they depend on only three parameters, and so (\ref{weyl}) does not represent the whole Weyl family but only those solutions with that particular value of their set of coefficients $a^W_n$. 

Since (\ref{unasigma}) is a solution of the linear equation (\ref{C2}) for $\sigma$ , the following expression is also a solution
\begin{equation}
\sigma(u,\bar u)=\sum_{k=0}^{\infty}\frac{2c_k}{\sqrt{(a_ku+2b_k)(a_k\bar u-2b_k)}}\label{sumadesigmas} ,
\end{equation}
and hence, the corresponding Weyl coefficients are given by a set of arbitrary parameters:
\begin{equation}
a^W_n =\sum_{k=0}^{\infty}\frac{2c_k}{a_k}\left(\frac{i2b_k}{a_k}\right)^n =\sum_{k=0}^{\infty}B_k^nC_k \ , \label{algebracond}
\end{equation}
with $B_k\equiv i2b_k/a_k$ and $C_k\equiv 2c_k/a_k$. And therefore, we generate another solutions by means of this linear combination.

In order to make a complete and detailed analysis of the results, we are going to show now that the solution previously obtained (\ref{sigma2}),(\ref{beta1}) belongs to the Weyl family for any analytic functions $v(\rho,z)$ and $w(\rho,z)$ considered, by means of performing explicitly a change of coordinates. 

It can be seen  that  a change of coordinates exists that allow us to recover the Weyl family of solutions from our solution. The change of coordinates is given by the following choice:
\begin{equation}
\hat \rho=\rho+2 v(\rho,z) \ ,
\label{rotilde}
\end{equation}
$v(\rho,z)$ being any analytic function, and the coordinate $\hat z$ must be  a solution of the Cauchy-Riemman equations: $\hat z_z=-\hat\rho_{\rho}$, $\hat z_{\rho}=\hat \rho_z$, and therefore both $v(\rho,z)$ and $w(\rho,z)$ satisfying the Cauchy-Newman problem (\ref{cauchynewman}) can be used to define the change of coordinates, i.e.,
\begin{equation}
\hat z=-\left(z+2w(\rho,z)+\mu\right) \ ,
\label{zetatilde}
\end{equation}
for an arbitrary constant $\mu$.

By substituting (\ref{rotilde}), (\ref{zetatilde}) in (\ref{sigma2}) we conclude that
\begin{equation}
\sigma(\hat\rho,\hat z)=\displaystyle{\frac{2c/a}{\sqrt{\hat\rho^2+(\hat z+\mu)^2+4pi(\hat z+\mu)-4p^2}}}\label{esta} \ ,
\end{equation}
where $p\equiv b/a$ and the previous expression (\ref{esta}) can be expanded as follows:
\begin{equation}
\sigma(\hat \rho,\hat z)=(2c/a) \sum_{n=0}^{\infty} \left[-(\mu+2pi)\right]^n \frac{P_n(\hat\Omega)}{\hat r^{n+1}} \ ,
\end{equation}
where $\hat \Omega \equiv \hat z/\hat r$, $\hat r$ being the  Weyl radial coordinate.

\section{Conclusion}

We have found  a solution of the static and axisymmetric vacuum field equations in a system of coordinates different from the canonical Weyl coordinates. We recall that the use of non-isotropic coordinates implies the determination of three metric functions, in contrast with the Weyl representation whereby one of these metric functions is chosen a priori. We write explicitly in some system of coordinates all the metric functions and we prove that the asymptotical conditions of the solution are satisfied. The family of solutions obtained depends on an analytic function that is a solution of certain Cauchy-Newman problem. It is proved that the canonical expression of the solution in Weyl coordinates is obtained  when the third metric function $J$ is equal to the standard Weyl form. 

The solutions obtained are not  new, in the sense that they belong to the Weyl family. The advantage of its present form is that they do not depend on a set of Weyl coefficients. From this solution we can construct some other family of  static and axisymmetric solutions by making linear combinations.

There exists a relationship between the gauge used to obtain the solution and the explicit form of the metric functions. The freedom of using different analytic functions in this generalization of coordinates may be used to look for the specific representation that make possible to write the Pure Multipole Solutions in a compact form.

\section{Acknowledgments}
This  work  was partially supported by the Spanish  Ministerio de Ciencia e Innovaci\'on
 under Research Project No. FIS 2009-07238, and the Consejer\'\i a de Educaci\'on of the Junta de Castilla y
Le\'on under the Research Project Grupo de Excelencia GR234.

We also wish to thank the support of the Fundaci\'on Samuel Sol\'orzano Barruso (University of Salamanca) with the project
FS/8-2009.

\end{document}